\documentclass[a4paper,12pt]{article}

\usepackage[utf8]{inputenc} %

\usepackage{appendix}
\usepackage{subcaption}
\usepackage{xcolor}
\usepackage{graphicx}
\usepackage{hyperref}
\usepackage[nolist,nohyperlinks]{acronym}
\usepackage{authblk}
\begin{acronym}
\acrodef{GAEN}{Google-Apple Exposure Notification}
\acrodef{TWR}{Two-Way Ranging}
\acrodef{BLE}{Bluetooth Low Energy}
\acrodef{UWB}{Ultra-Wide Band}
\acrodef{BLEnd}{Bluetooth Low Energy Neighbor Discovery}
\acrodef{FBK}{Fondazione Bruno Kessler}
\acrodef{RSSI}{Received Signal Strength Indicator}
\end{acronym}

\definecolor{LowRisk}{HTML}{43978D}
\definecolor{MediumLowRisk}{HTML}{F9E07F}
\definecolor{MediumHighRisk}{HTML}{F9AD6A}
\definecolor{HighRisk}{HTML}{D46C4E}

\usepackage{xspace}
\newcommand{\idandalo}{\textsc{am-pri}\xspace}
\newcommand{\idpovokids}{\textsc{day-pri}\xspace}
\newcommand{\idpovoteens}{\textsc{day-int}\xspace}

\title{Measuring close proximity interactions in summer camps during the COVID-19 pandemic}
\date{\today}

\author[1,2]{E. Leoni\thanks{eleoni@fbk.eu}}
\author[1]{G. Cencetti}
\author[1]{G. Santin}
\author[3]{T. Istomin}
\author[3]{D. Molteni}
\author[3]{G. P. Picco}
\author[1]{E. Farella}
\author[1]{B. Lepri}
\author[1]{A. M. Murphy}

\affil[1]{DIGIS, Bruno Kessler Foundation, Trento, Italy}
\affil[2]{DEI, Universit{\`a} di Bologna, Bologna, Italy}
\affil[3]{DISI, Universit{\`a} di Trento, Trento, Italy}

\begin{document}
\maketitle

\begin{abstract} 
Policy makers have implemented multiple non-pharmaceutical strategies to mitigate the COVID-19 worldwide crisis. 
Interventions had the aim of reducing close proximity interactions, which drive the spread of the disease. A deeper knowledge of human physical interactions has revealed necessary, especially in all settings involving children, whose education and gathering activities should be preserved. Despite their relevance, almost no data are available on close proximity contacts among children in schools or other educational settings during the pandemic.

Contact data are usually gathered via Bluetooth, which nonetheless 
offers a low temporal and spatial resolution. Recently, ultra-wideband
(UWB) radios emerged as a more accurate alternative that nonetheless
exhibits a significantly higher energy consumption, limiting in-field
studies. In this paper, we leverage a novel approach, embodied by the
Janus system that combines these radios by exploiting their
complementary benefits. The very accurate proximity data gathered
in-field by Janus, once augmented with several metadata, unlocks
unprecedented levels of information, enabling the development of novel
multi-level risk analyses.

By means of this technology, we have collected real contact data of
children and educators in three summer camps during summer 2020 in the
province of Trento, Italy.  The wide variety of performed daily
activities induced multiple individual behaviors, allowing a rich
investigation of social environments from the contagion risk
perspective. We consider risk based on duration and proximity of
contacts and classify interactions according to different risk
levels. We can then evaluate the summer camps' organization, observe
the effect of partition in small groups, or social bubbles, and
identify the organized activities that mitigate the riskier
behaviors. 

Overall, we offer an insight into the educator-child and
child-child social interactions during the pandemic, thus
providing a valuable tool for schools, summer camps, and policy makers
to (re)structure educational activities safely.


\end{abstract}



\section{Introduction}
Close proximity interactions (CPIs) drive the spread of any disease that is transmitted predominantly by respiratory droplets and saliva, such as influenza, common colds, and severe acute respiratory syndromes (i.e., SARS, MERS, COVID-19)~\cite{brankston2007,read2008,funk2010,salathe2010,huang2016,leung2021}. An improved characterization of CPIs should thus lead to a better understanding of the spread dynamics and possibly inform public health experts and policy makers to design more effective interventions~\cite{weinstein2003}.

For this reason, some research efforts have used wearable devices and Radio Frequency Identification (RFID) or Infrared (IR) sensors to measure and analyze high-resolution proximity interactions in different settings such as schools~\cite{salathe2010,stehle2011s}, workplaces~\cite{cattuto2010,alshamsi2015}, hospitals~\cite{isella2011,vanhems2013,hertzberg2017,duval2018,duval2019}, households~\cite{ozella2018}, and conferences~\cite{cattuto2010,isella2011j,stehle2011}.

During the COVID-19 pandemic, social contacts and in particular CPIs were significantly modified~\cite{jarvis2020,zhang2020,feehan2021quantifying,zhang2021} by several non-pharmaceutical interventions such as physical distancing measures (i.e., 1~m or more), mobility restrictions, closings of schools, universities, and selected businesses (e.g., restaurants, bars, coffee shops, gyms), promotion of teleworking,
cancellations or limits on the size of events (e.g., sports events, weddings, funerals), limits on
the number of people in small family, educational and social gatherings (i.e., social bubbles), etc.~\cite{haug2020ranking,hsiang2020,brauner2021}.


However, despite their relevance, almost no data are available on how CPIs occur among children in contexts such as schools or summer camps during the COVID-19 pandemic, thus making it difficult to evaluate and model the effects of physical distancing measures, small group strategies, preferences for outdoor activities, masks, etc., on CPIs, as well as identifying the situations and activities during school and summer camp days where the risk of transmission is elevated.

The collection of reliable data in these environments (e.g., schools, summer camps) is itself a nontrivial task. During the pandemic, several local and national governments have launched smartphone digital contact tracing (DCT) apps based on the Bluetooth Low Energy (BLE) technology~\cite{gomez2012} and
the GAEN (Google and Apple Exposure Notification) interface~\cite{GAEN}, and  several studies have shown the effectiveness of Bluetooth-based DCT using real-world contact patterns~\cite{cencetti2021,barrat2021} and in pilot and country-wide studies conducted in Switzerland, the United Kingdom (the
Isle of Wight and the whole country), and Spain (Gomera island)~\cite{salathe2020,kendall2020,rodriguez2021, wymant2021}.

In addition to the challenge that most children do not carry personal
smartphones, this technology has at least two shortcomings for
capturing CPIs in schools and summer camps: (i) \emph{low temporal
  resolution} (i.e., GAEN detects neighbors every
4~minutes~\cite{GAEN}), and (ii) \emph{low spatial resolution}, which
directly descends from limitations of BLE and leads to significant
estimation errors~\cite{leith20:plosone}. The first issue can be
tackled by the use of an alternative to GAEN, while the second can be
addressed by changing the technology used for estimating distances,
e.g., to ultra-wideband (UWB), which brings the spatial error down
from meters to decimeters~\cite{zafari_survey_2019}.

In this paper, we address these issues via a novel approach, embodied
in the Janus system~\cite{istomin2021janus}, combining a custom,
efficient device discovery mechanism based on BLE with the ability to
accurately measure pairwise distances via UWB. In our experiments, we
configured Janus to acquire distance measurements every 30~s and
installed it on a wearable device that children can easily carry.
We have collected real-world CPIs with Janus at three summer
camps in the province of Trento (Italy). These camps offer interesting
settings because of the rich variety of daily activities that induce
different CPIs among children and between children and the summer
camps' educators. Moreover, the summer camps took place during the
summer of 2020, in the middle of the pandemic and just after the local
easing of lockdown measures. As such, it is possible to investigate
the effect of the guidelines and regulations enforcing physical
distancing, mask-wearing, outdoor activities, and the formation of
small groups (i.e., social bubbles).

The accurate and fine-grained contact data uniquely enabled by Janus,
complemented by the metadata about summer camps, results in the rich
data set that is the basis of our multi-level analysis. First, we
explore the definition of \emph{close contact} as the aggregation of
multiple raw measurements captured by the sensors and discuss the
modeling choices implied by this operation.  After this aggregation
phase, the resulting contacts are enriched with metadata. For example,
social bubbles~\cite{block2020social,leng2020effectiveness} were
enforced as a contagion containment measure, and thus we assign to
each contact the groups of the two involved individuals. Further, each
contact is associated with the activity being performed during the
contact time.

By considering the metadata in the analysis along with the raw contact
data, we offer novel insights into both educator-child and child-child
social interactions during the pandemic. In particular, we study the
distribution of the level of contagion risk among individuals
depending on the proximity and duration of their contacts, finding
that a vast majority of CPIs are classified as low risk. Moreover, we
aggregate the contacts as intra-group (i.e., within the social bubble)
and inter-group (i.e., between different bubbles), and observe changes
in the distribution of contact risk levels in the two cases, offering
evidence of the effectiveness of the social bubble strategy
\cite{block2020social,leng2020effectiveness}. Finally, a thorough
analysis of the different activities provides insights into their
inherent risks of contagion, which can be further interpreted in view
of the features of the activity itself (indoor or outdoor, static or
dynamic, etc.).

The results of our analyses provide information immediately actionable by school and summer camp managers and teachers, policy makers, and public health experts.

\section{Materials and Methods}

We concisely describe the salient aspects of the Janus system used in
our in-field experiments, offer details about the summer camps where
they were performed and the mechanics of data acquisition, and state
the definition of close proximity contact used throughout the paper.

\subsection{Janus: A system for measuring close proximity interactions}

Janus~\cite{istomin2021janus} relies on a dual-radio architecture to
provide an accurate and energy-efficient system for proximity
detection.  We split proximity detection into two primary
functionalities: identifying the other devices nearby and measuring
the distances between them.

The first, device discovery, must be performed continuously as people
(and the devices they carry) move freely in an unconstrained
space. Fundamentally, Janus detects that two devices are near each
other when they are able to communicate. For this continuous
operation, we exploit the lower power BLE radio and build atop the
BLEnd~\cite{julien17:ipsn} continuous neighbor discovery
protocol. BLEnd defines the optimal schedules for the BLE
advertisement and scan periods to minimize consumption while meeting a
service level agreement defined by the maximum allowed latency to
discovery, the required probability for discovery, and the maximum
number of devices expected to be in range. In our experiments, we
configured the BLEnd component of Janus to guarantee the discovery of a
neighbor within 30~s at least 95\% of the times, provided no more than
20~devices are in range.

Once a nearby device is detected, Janus exploits the payload of the
BLE advertisements continuously sent by BLEnd to coordinate, at no
additional communication cost, the accurate ranging between devices
performed by the UWB radio. Janus relies on single-sided two-way
ranging (SS-TWR), part of the IEEE~802.15.4
standard~\cite{std154}. This scheme requires a 2-packet exchange
between an initiator and a responder; the transmission and reception
of these packets are timestamped and made available at the initiator,
which can compute the time of flight and therefore the distance
between devices. In Janus, each device periodically schedules a
ranging window during which it is available to respond to ranging
requests; the aforementioned coordination exploiting BLE advertisements
informs each neighbor of its unique offset into this window, ensuring
that ranging requests among multiple neighbors do not
collide. Additional details about Janus are available
in~\cite{istomin2021janus}.

\begin{figure}[!t]
    \centering
    \captionsetup{type=figure}
    \includegraphics[width=.9\textwidth]{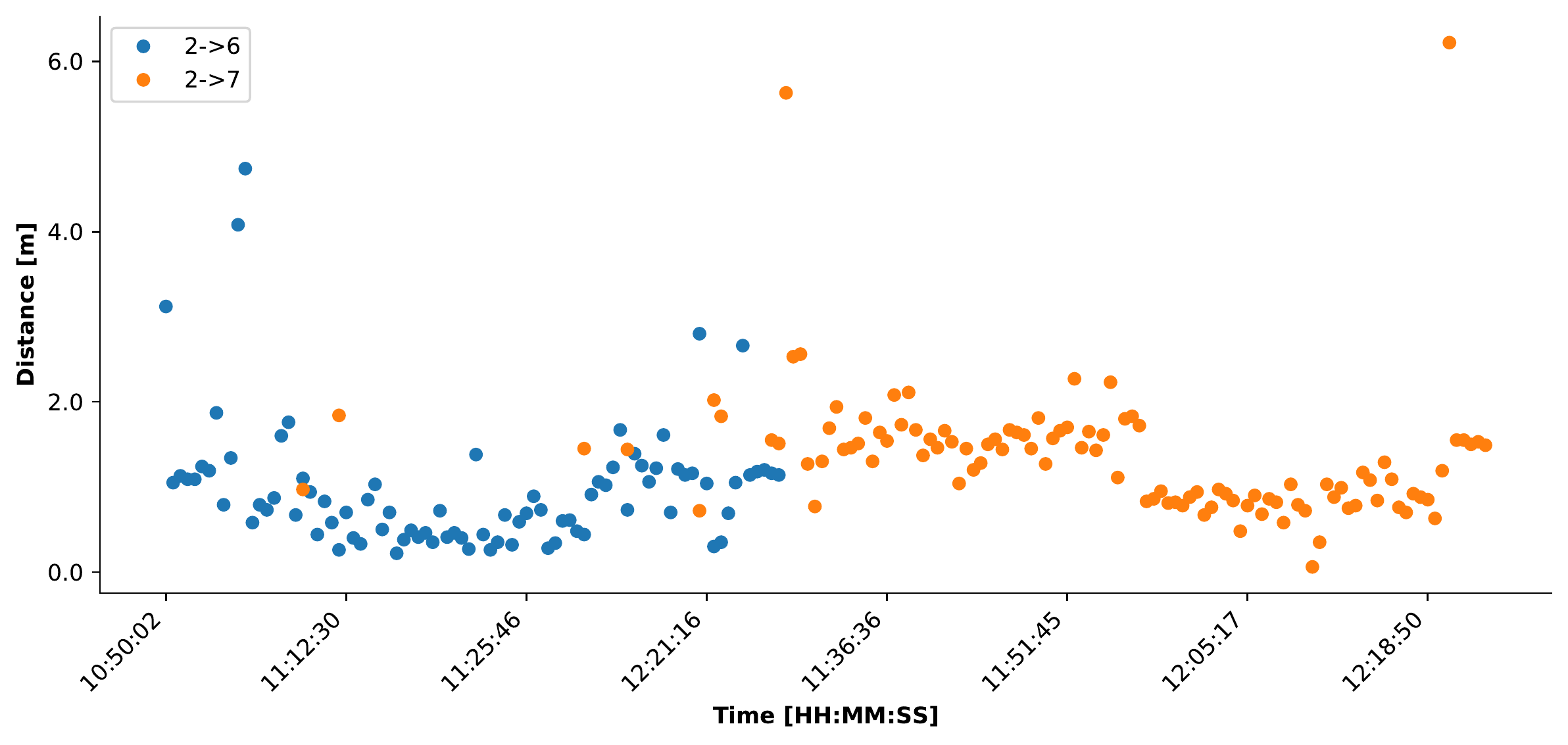}
    \caption{ {Raw data obtained from a Janus device.}
      Timestamped distance measurements collected over 105~min by the
      Janus device of user~2 with respect to the devices of user~6 and~7. }
    \label{fig:scatter_plot}
\end{figure}

Figure~\ref{fig:scatter_plot} provides an example of the distance data
directly obtained from Janus devices, which in our case are the
MDEK1001 development kits by Decawave (now Qorvo), equipped with a BLE
radio and the popular DW1000 UWB transceiver. The chart shows a
snapshot of 105~minutes for device~2 with respect to two other
devices, 6~and~7. Each dot indicates the distance measurement between
device~2 and either device, color-coded as blue for device~6 and
orange for device~7. According to our configuration, samples are taken
every 30~s. Even without additional data processing, it can be easily
seen that device~2 (and therefore the person carrying it) was very
close (within 1~m) to device~ 6 for approximately 15~min, starting
just after 11:00. We also note that the data is quite \emph{clean};
the variations of the measurements across time are consistent. This is
due to the accuracy of UWB, which enables our subsequent analysis.

\subsection{Data acquisition}\label{sec:data_description}

\begin{table}[!t]
\caption{Description of the three summer camps investigated in our study.}
\label{tab:experiments}
\begin{tabular}{p{12mm}p{40mm}p{10mm}p{10mm}p{10mm}p{10mm}}
\hline
ID & Short Description & Ages& Children& Educators&Groups\\
\hline
\idandalo & Morning camp with a large indoor space, nearby a public park.&6-11 & 21 & 5 & 3\\ 
\idpovokids & All day  camp in an alpine region with only outdoor space. & 6-11 & 13 & 5 & 2\\ 
\idpovoteens & All day camp in an alpine region with additional indoor space. &11-14 & 9 & 2 & 1\\ 
\hline
\end{tabular}
\end{table}

\begin{table}[!t]
  \caption{Daily activities at the summer camps, each with a brief description, the location and the duration in minutes for each summer camp that offered the 
activity.
}
\label{tab:activities}
  \begin{tabular}{p{19mm}p{33mm}p{12mm}p{11mm}p{11mm}p{11mm}}\hline
  Activity & Description & Location & \idandalo&\idpovokids&\idpovoteens \\\hline
  Woods & Playing in a wooded area & outdoor & 90~min& & \\
  Soccer & Playing in a soccer field & outdoor & 90~min& &\\
  Board games & Playing tabletop games & indoor & 90~min \\
  Newspaper & Pairs work at computers & indoor & 90~min& \\
  Theater & Singing and acting & indoor & 90~min&& \\
  Snack & Short food break  & indoor & 15~min && \\
  Team games & Organized group games & indoor & 90~min& 120~min&120~min \\
  Crafts & Arts and craft & indoor & 90~min&& 180~min  \\
  Hiking & Group walk &  outdoor & &&240~min  \\
  Round table & Greetings, planning, etc &   indoor & &&180~min \\
  Day closing & Free play pre pick-up  & outdoor & &30~min&60~min \\
  Outdoor lunch & Eating & outdoor & &&60~min \\
  Indoor lunch & Eating & indoor & &60~min& \\
  Free play & No organized activities & indoor & &&60~min \\
  Free play & No organized activities & outdoor &&60~min& \\
  \hline
  \end{tabular}
  
\end{table}

The data used in our analyses results from a study conducted from August to September 2020 in three different summer camps, summarized in 
Table~\ref{tab:experiments}, in Trentino, Italy.
The study design was approved by the Agency for Family, Birth, and Youth Policies (Agenzia Provinciale per la Famiglia, la Natalit\`a, e le Politiche Giovanili) 
of the Autonomous Province of Trento\footnote{\href{https://www.trentinofamiglia.it/}{https://www.trentinofamiglia.it/}}, the provincial government body 
responsible for the organization of the summer camp programs, and by the two social cooperatives directly responsible for camp management and activities. 
Kaleidoscopio\footnote{\href{https://www.kaleidoscopio.coop/}{https://www.kaleidoscopio.coop/}}, the two social cooperatives responsible of the daily management 
of the two summer camps where the study took place.
In preparation for the study, parents and educators were provided with detailed information about the purpose of the study, the data treatment and privacy 
enforcement strategies, the devices the children and educators would be using, and the measurements they provide. Following Italian regulations, all parents and 
educators signed an informed consent form. Special attention was given to privacy and data protection: no personal information was associated with the 
identifier of the corresponding Janus device. We did note the group (i.e., social bubble) the individual belonged to and, in some cases, the identity of devices 
carried by others for whom physical distancing rules were waived (e.g., among siblings and between children with special needs and the educators assigned to 
assist them).

The first summer camp, \idandalo, operated for half days (mornings) with 21 primary school-age children and 5 adult educators, all of whom agreed to participate 
in the study.
The children were divided into 3 groups, each with one or two educators. Each activity during the day was restricted to a single group at a time to maintain 
separation and leverage the concept of social bubbles~\cite{block2020social,leng2020effectiveness}.

The second and third camps were organized the same week by the same cooperative, but took place at different locations; therefore, we treat them separately.
Both were all-day camps from 8:00 to 16:30. \idpovokids applied the social bubble with two groups of primary school children. The third camp, \idpovoteens, 
involved 9 intermediate school children with two educators. 
The overall participation rate in these two camps was 94\%.

The summer camps engaged the children in different educational and playing activities, as summarized in Table~\ref{tab:activities}. For each activity, we 
indicate the approximate duration in minutes for each camp.

\subsubsection{Device setup and experimental setting}
\begin{figure}
    \centering
    \includegraphics[width=.28\textwidth]{ 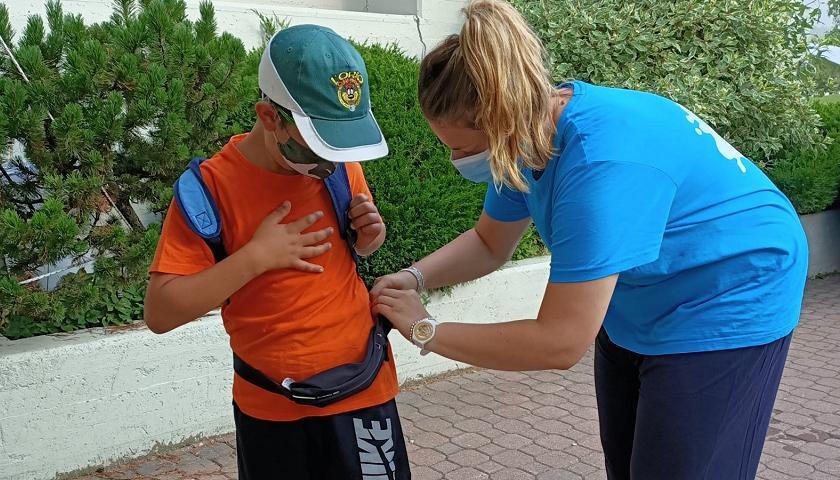}
    \includegraphics[width=.67\textwidth]{ 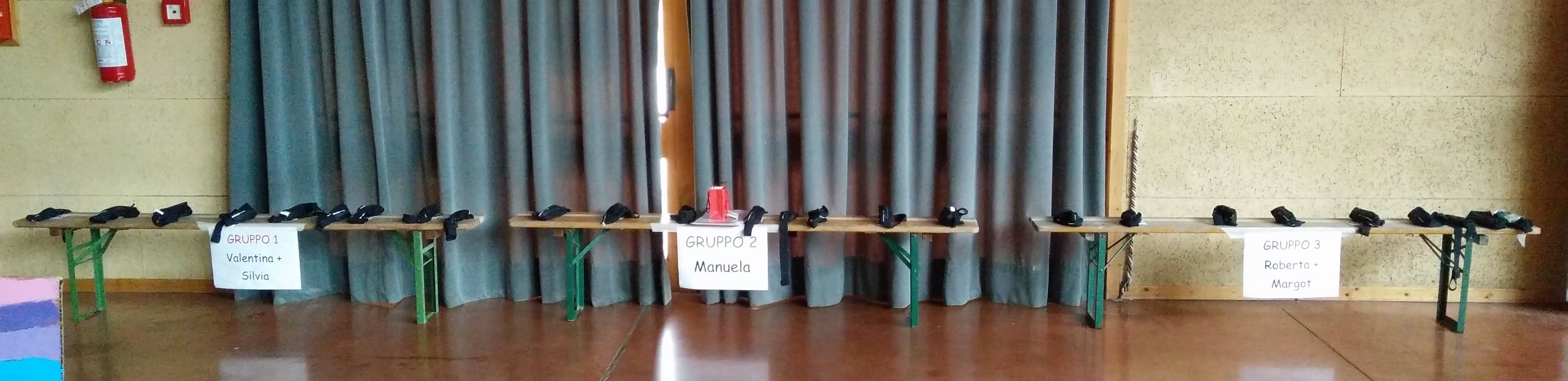}
    \caption{ {Janus device management at the} \idandalo
       { camp.} Left: An educator fitting the waist bag
      containing the device on a child, on the first camp day. Right:
      Devices in waist bags sitting on a storage bench overnight; the
      inhibitor device is inside the red bag in the center. }
    \label{fig:andalo}
\end{figure}

To make carrying the device comfortable for the children, we inserted
it inside a waterproof waist bag, as shown on the left of
Figure~\ref{fig:andalo}. We received positive feedback from the
educators, who said that the children immediately forgot they were
wearing the device.  As mentioned, the Janus device is configured to
sample distances every 30~s when devices are in proximity. Measurements
greater than 10~m are discarded to save memory on the device and
because these large distances are not considered relevant for the
transmission of SARS-CoV-2~\cite{jones2020two,chu2020physical}.

After programming the devices and inserting new batteries, the waist
bags were delivered to camp organizers at the beginning of each
week. The educators were responsible for handing out the bags to the
same children each morning and collecting them at the end of the
day. At the end of the week, the devices were collected and the data
offloaded via USB.

As the devices do not have an on/off switch, to avoid the collection
of meaningless data at night, when devices were stored on a bench
(Figure~\ref{fig:andalo}), we implemented an \emph{inhibitor} device.
This special device was turned on at the end of the day by connecting
it to a USB power bank. When the regular devices detected the BLE
advertisement of the inhibitor, they went to sleep for 5~min. Upon
restarting, if the inhibitor was detected again, they returned to
sleep; otherwise, they started functioning normally, ranging with all
neighboring devices. Each morning, the inhibitor device was detached
from its power supply. This inhibition mechanism saved battery as well
as memory and, most important, required no technical skills from the
educators; even using the USB power bank was much easier than removing
the battery from all Janus devices, which was the only other
alternative available.

Thus, since the devices have not an on/off button, to avoid after-hours data (i.e., during night) we implemented an inhibitor node. This special device was 
attached to a USB power bank and it continuously sent BLE packets which told the other nodes to not do ranging operations between them but just stop each 
operation for 5~minutes and than check again whether they received a packet from that special device. In case not, the node restarted to work normally. The 
advantages of this mechanism were threefold: (i) the battery level was almost maintained from one day to another, (ii) the memory was not filled with 
meaningless data (the pouches were all located nearby during the night), and (iii) the educators did not have to open each day the devices in order to remove 
the batteries and insert them again the next day.

\subsection{Definition of close proximity contacts}\label{sec:contact_def}


After downloading the measurements from all devices, we pre-processed
them as detailed in Appendix~\ref{si:pre_processing}. This processing
removed spurious measurements, e.g., those recorded between the
morning activation of the devices and the start time of the
activities. We then aggregated these raw samples into \emph{contacts}
characterized by two device IDs, the timestamp marking the beginning
of the contact, the contact duration, and a distance, as described
next.

To identify a contact, we focus on a pair of IDs, collecting all
measurements collected by either device, and sorting them in
time. This sequence is then processed sequentially to divide the time
into multiple, meaningful contacts. Intuitively, a contact should
contain measurements that are all \emph{temporally and spatially
  close} to one another, which we define via time and distance
thresholds.

We begin with the temporal dimension, splitting the sequence into
sub-sequences whenever a gap of $\tau_{time}=90~s$ exists between two
consecutive measurements. This step accounts for interruptions in the
interaction between the pair of devices, e.g., when they move away
from one another.

Second, we check each of the distances inside each sub-sequence,
ensuring that a single contact contains only measurements with similar
distances, and ensuring that the single distance attribute assigned to
a contact has a reasonable spatial variation.  Therefore, we
sequentially process the measurements of a sub-sequence in temporal
order, and retain them in a single sub-sequence as long as all the
measured distances are within $\tau_{space}=2 ~m$ from each
other;
a new sub-sequence is started upon the first measurement outside this range.

In this way, we obtain a set of sub-sequences, each containing
measurements without large temporal gaps and with similar
distances. After discarding sub-sequences with fewer than
$\tau_{len}=2$ measurements, we aggregate each cluster into a
contact. Each contact is tagged with the timestamp of the first
measurement in the sub-sequence, a duration given by the time span of
the measurements in it, and a distance given by the median value of
the measurements. Using the median (i.e., the central value of the
distribution) yields a more robust value compared to the mean, which
is more sensitive to extreme values and outliers.

\begin{figure}[!t]
    \centering
    \captionsetup{type=figure}
    \includegraphics[width=.7\textwidth]{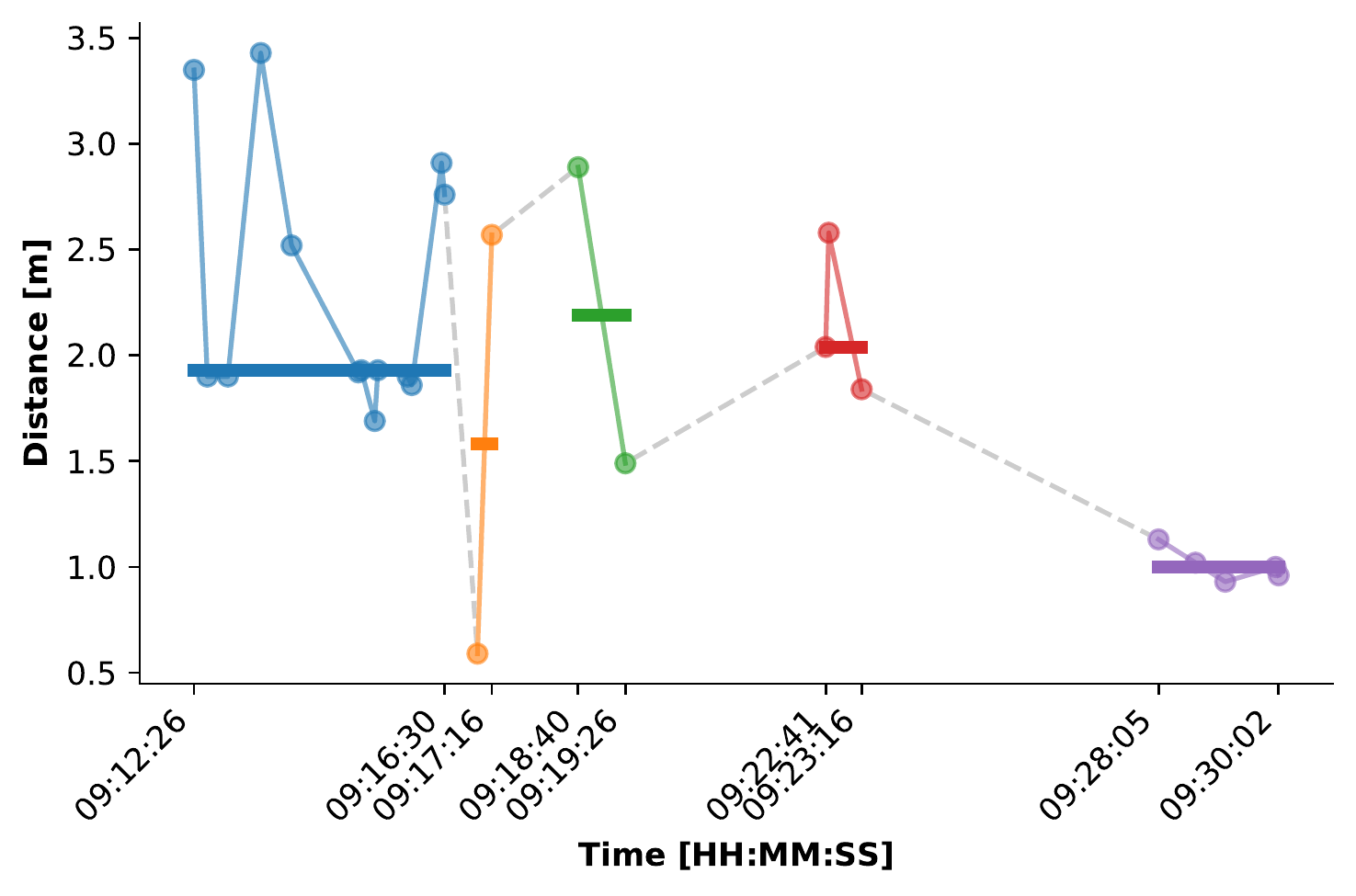}
    \caption{ {Measurement splitting and contact aggregation
        process.} The figure shows the measurements collected in the
      first 20~min of August 8th, 2020, between node $26$ and $27$ at
      the \idandalo camp. The measurements (light colors) are colored
      according to the division into contact characterized by
      $\tau_{time} =90 ~s$ and $\tau_{space} =2~m$. Each contact is
      depicted as a horizontal bar from its beginning to its end,
      where the height of the bar represents the median distance.}
    \label{fig:contact_formation_example}
\end{figure}

An example of this splitting and aggregation process is shown in Figure~\ref{fig:contact_formation_example}, which depicts a sequence of measurements in a 
20~min period grouped into sub-sequences (identified by colors) and aggregated into contacts (identified by the horizontal lines). The different splitting 
strategies can be observed. For example, the orange and green sequences are separated due to the gap of more than $\tau_{time}$ between them. On the other hand, 
the blue and orange sequences are separated because the first measurement in the orange cluster is outside the range of $\tau_{space}$ with respect to the 
previous measurements.

The resulting contacts model the high-level notion of CPI that we use
in our analyses in the next sections, and enables the general
contagion risk assessment of the different environments. Further, we
also associate to each contact the groups of the involved IDs and an
activity when both IDs are in the same group.

Some of the contacts can be removed a posteriori to account for
risk-modelling choices. For instance, we discard contacts between
siblings (who were not required to respect physical distancing rules)
or between children with special needs and their support teacher.
Additionally, in \idpovokids and \idpovoteens, the two activities
``welcoming activity'' and ``swimming pool'' have been discarded
because the devices had not all been distributed and were piled up in
the same place, resulting in many spurious measurements.

The resulting numbers of contacts for each summer camp setting are reported in Table~\ref{tab:datasets_cleaned}. For each data set, we also report the number 
and percentage of contacts where both users belong to the same group, and thus to which we are able to assign an activity.

\begin{table}[!t]
\caption{Description of the contacts resulting from the aggregation procedure. For each camp, we report the total number of contacts, the average number of the 
measurements for each contact, the number of groups and activities in the camp, and the number and percentage of the contacts that are uniquely associated with 
an activity. For \idpovokids and \idpovoteens, we report both the number of activities, and the number of activities considered for the analysis (in 
parenthesis).}
\label{tab:datasets_cleaned}
\begin{tabular}{p{20mm}p{12mm}p{15mm}p{15mm}p{15mm}p{23mm}}
\hline
ID & Num. Contacts & Average Measurements per Contact & Num. Groups & Num. Activities &  Activity-tagged contacts \\
\hline
\idandalo   & 7259 & \phantom{1}5.80 &  3 &  8 & 6833 ( 94.13 \%)\\ 
\idpovokids  & 7561 & \phantom{1}8.48 &  2 &  5 (4) & 6774 ( 89.59 \%)\\ 
\idpovoteens  & 3485 & 16.40 &  1 &  9 (7) & 3485 (100.00 \%)\\
\hline
\end{tabular}
\end{table}

\section{Results}
Leveraging the previous definition of contacts and additional metadata, we can now delve into the analysis of the complex daily CPI patterns within the summer 
camps.

\subsection{Identification of contagion risk levels} 

To build a general model for risk analysis, we define four different categories of contagion risk for contacts based on proximity and duration. We then classify 
all contacts into these categories.

In a meta-analysis and systematic review of observational studies on
SARS-CoV, MERS-CoV, and SARS-CoV-2 person-to-person
transmission~\cite{chu2020physical}, a physical distancing of less
than 1~m was reported to result in a significantly higher transmission
risk than distances higher than 1~m (12.8\% vs. 2.6\%), thus supporting
a minimum physical distance of 1~m, as in the rule enforced in schools
and summer camps in Italy.  However, as pointed out by Jones et
al.~\cite{jones2020two}, physical distancing rules would be more
appropriate and effective if they offer graded levels of
risk. Similarly, although contact tracing guidelines in several
countries, various digital tracing contact apps, and some
studies~\cite{cheng2020} assume that the duration of exposure to a
person with COVID-19 influences the transmission risk (e.g., defining
a threshold of 15~min beyond which transmission risk increases), a
precise quantification of the duration of exposure is still
missing~\cite{jones2020two}.

Following these considerations, we define the risk categorization
summarized in Table~\ref{tab:risk}.
The first category is associated with a \emph{high risk} of contagion
and includes all contacts with duration above 15~min and distance less
than 1~m. The second category, \emph{medium-high risk}, includes all
contacts with duration above 10~min and distance below 2~m that are
not included in the high-risk category. The third category,
\emph{medium-low risk}, includes contacts with duration above 5~min
and distance below 4~m not included in the previous categories.  The
fourth category contains all remaining contacts, therefore associated
to a \emph{low risk} level.

Notably, this granularity in discriminating risk levels is enabled by
the fine-grained spatio-temporal resolution offered by Janus, and
would be simply unfeasible with the significantly coarser one offered
by BLE approaches. For instance, using the GAEN interface, a 5~min
window would contain a \emph{single} sample, let apart the inherent
inaccuracy of the corresponding BLE-based distance estimation.

\begin{table}[!t]
\caption{Risk levels of contagion defined on the basis of duration of exposure and physical distance. }
\label{tab:risk}
\centering
\begin{tabular}{lcc}
\hline
 & Duration & Distance \\
\hline
 {\Large{\textcolor{HighRisk}{$\bullet$}}} High risk & $\geq$ 15~min & $\leq$ 1 m\\
 {\Large{\textcolor{MediumHighRisk}{$\bullet$}}} Medium high risk & $\geq$ 10~min  & $\leq$ 2 m\\
 {\Large{\textcolor{MediumLowRisk}{$\bullet$}}} Medium low risk & $\geq$ \phantom{1}5~min & $\leq$ 4 m\\
 {\Large{\textcolor{LowRisk}{$\bullet$}}} Low risk & $<$ \phantom{1}5~min & $>$ 4~m\\
 \hline
\end{tabular}
\end{table}

\subsection{Contagion risk analysis}

\begin{figure}[!ht]
\centering
\captionsetup{type=figure}

\begin{subfigure}{.7\textwidth}
\includegraphics[width=1.\textwidth]{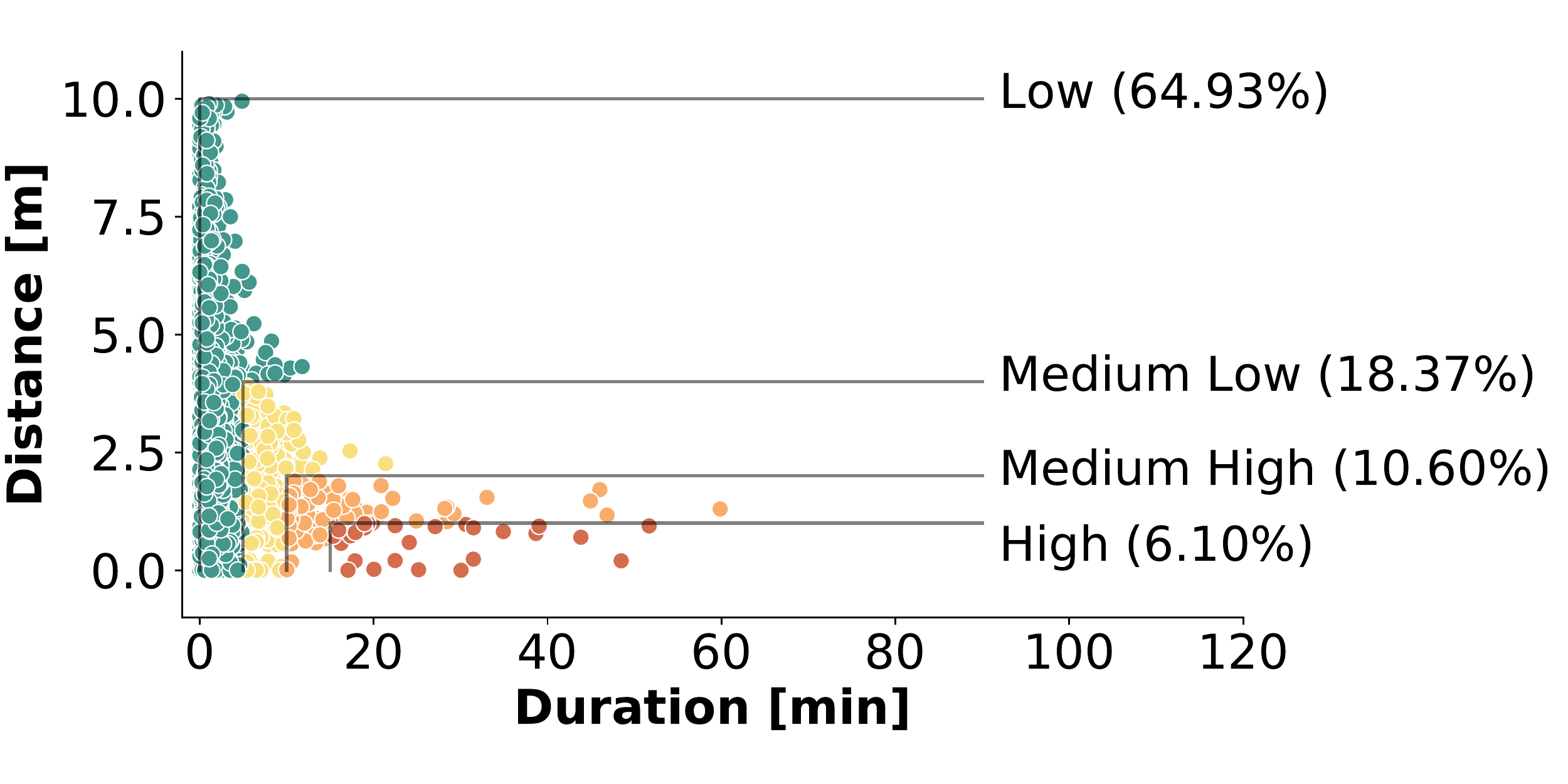}
\caption{\idandalo}\label{fig:risk_all_andalo}
\end{subfigure}

\begin{subfigure}{.7\textwidth}
\includegraphics[width=1.\textwidth]{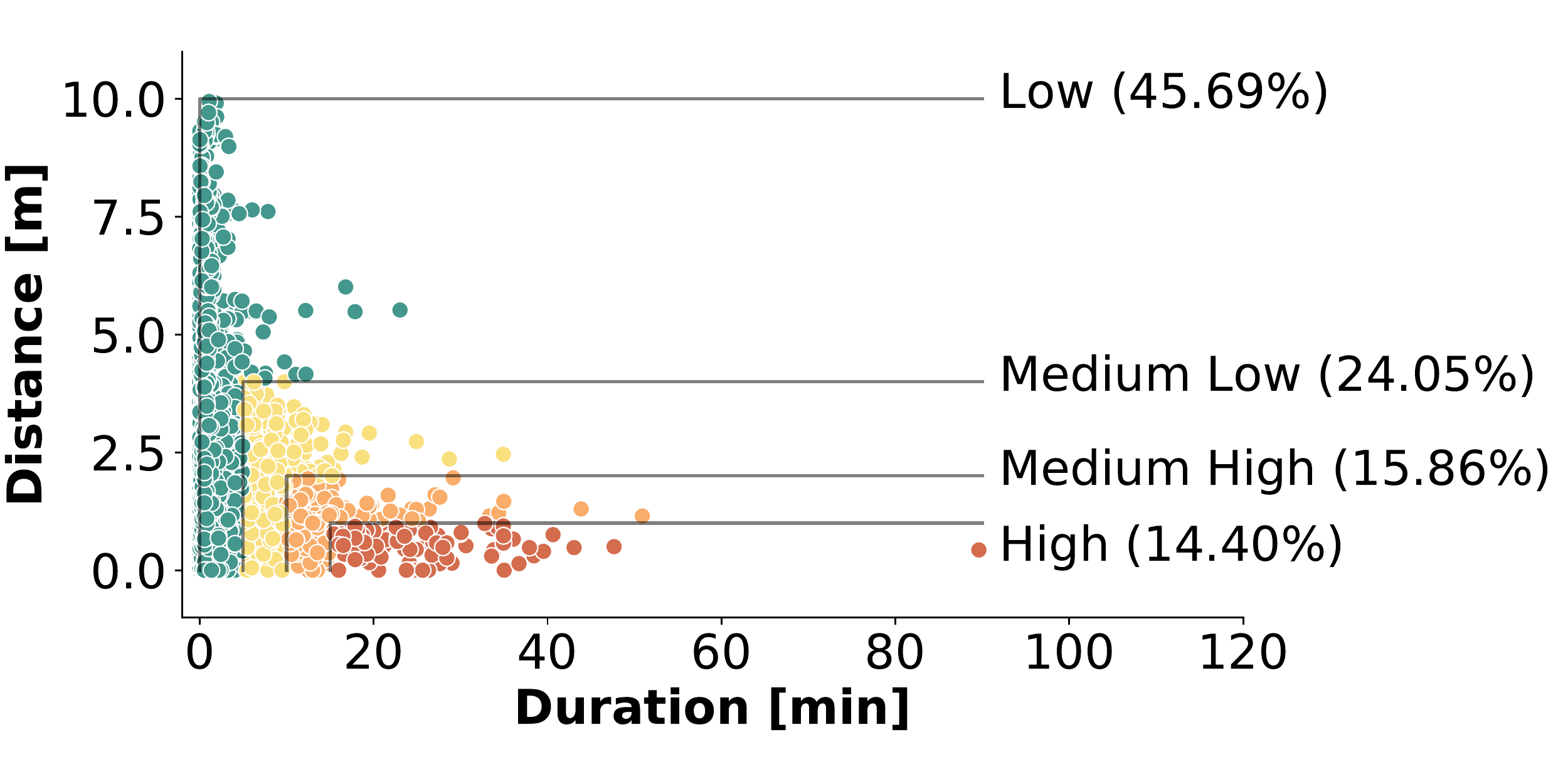}
\caption{\idpovokids}\label{fig:risk_all_povo_kids}
\end{subfigure}

\begin{subfigure}{.7\textwidth}
\includegraphics[width=1.\textwidth]{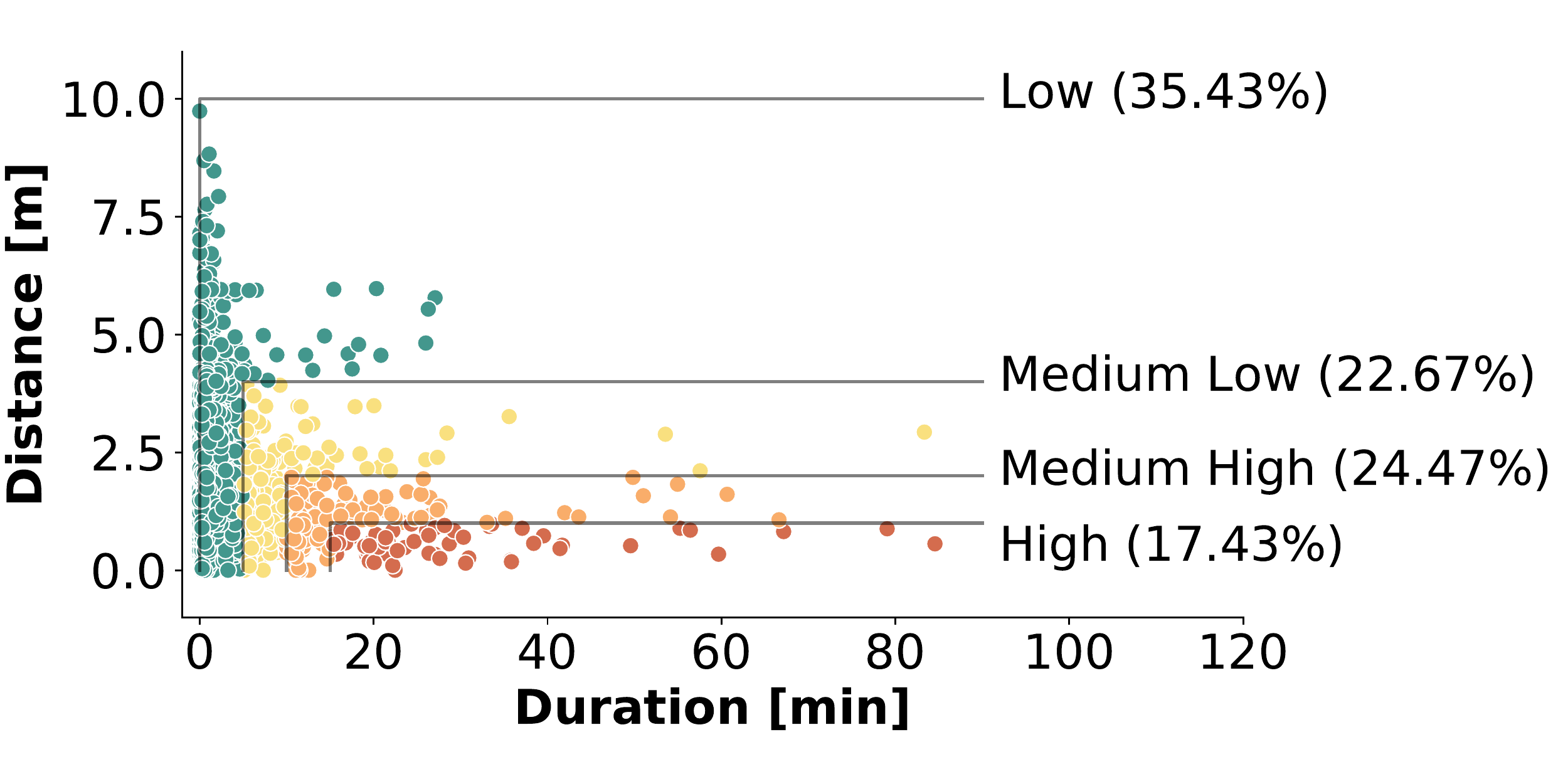}
\caption{\idpovoteens}\label{fig:risk_all_povo_teens}
\end{subfigure}

\caption{ {Summer camp contacts and contagion risk.} The
  figure reports, for each summer camp, the corresponding contacts
  classified according to their risk of contagion as a function of
  the duration of exposure and proximity, following the risk categories in
  Table~\ref{tab:risk}. The values in parentheses denote the
  percentage of time spent in a contact with the corresponding
  risk category.}
\label{fig:risk_all}
\end{figure}

Figure~\ref{fig:risk_all}
shows a scatter plot for each summer camp dataset, reporting the recorded contacts as a function of duration and proximity. Each dot represents a contact, as 
defined in Section~\ref{sec:contact_def}, with colors describing the associated risk according to the color code in Table~\ref{tab:risk}.
The percentages reported inside the figures, and associated with the different risk levels, represent the percentage of time spent by the population in the 
corresponding risk category. 
Interestingly, we see that, even if different summer camps imply different levels of risk, there is a non-negligible percentage of contacts at high risk of 
contagion in all summer camps.

\begin{figure}[!t]
\captionsetup{type=figure}
\centering
\begin{subfigure}{.7\textwidth}
\captionsetup{type=figure}
\centering
\includegraphics[width=\textwidth]{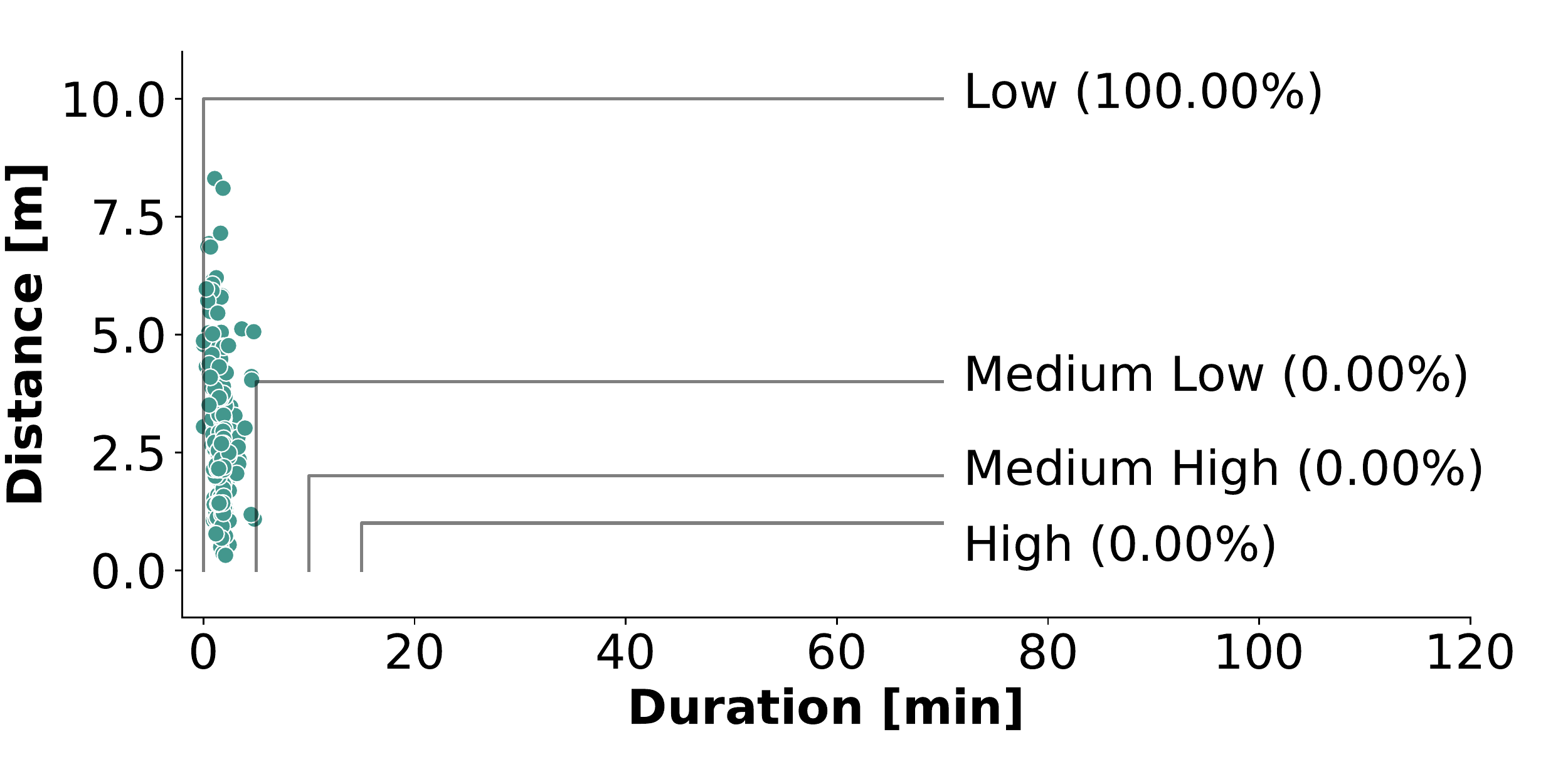}
\caption{
  Mean duration of exposure and mean spatial distance.}\label{fig:risk_unique_mean}
\end{subfigure}

\begin{subfigure}{.7\textwidth}
\captionsetup{type=figure}
\centering
\includegraphics[width=\textwidth]{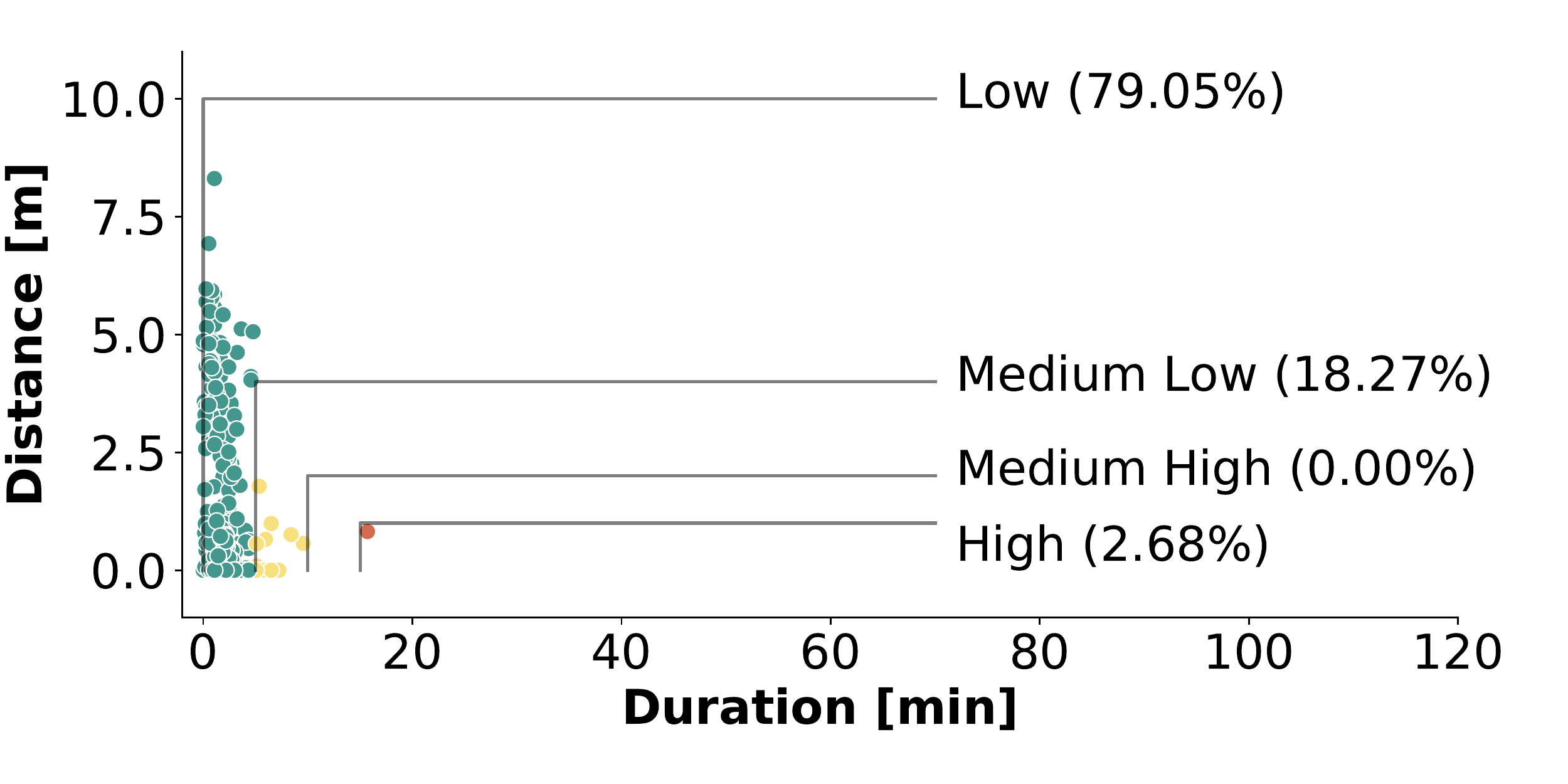}
\caption{
  Minimal spatial distance.}\label{fig:risk_unique_dist}
\end{subfigure}

\begin{subfigure}{.7\textwidth}
\captionsetup{type=figure}
\centering
\includegraphics[width=\textwidth]{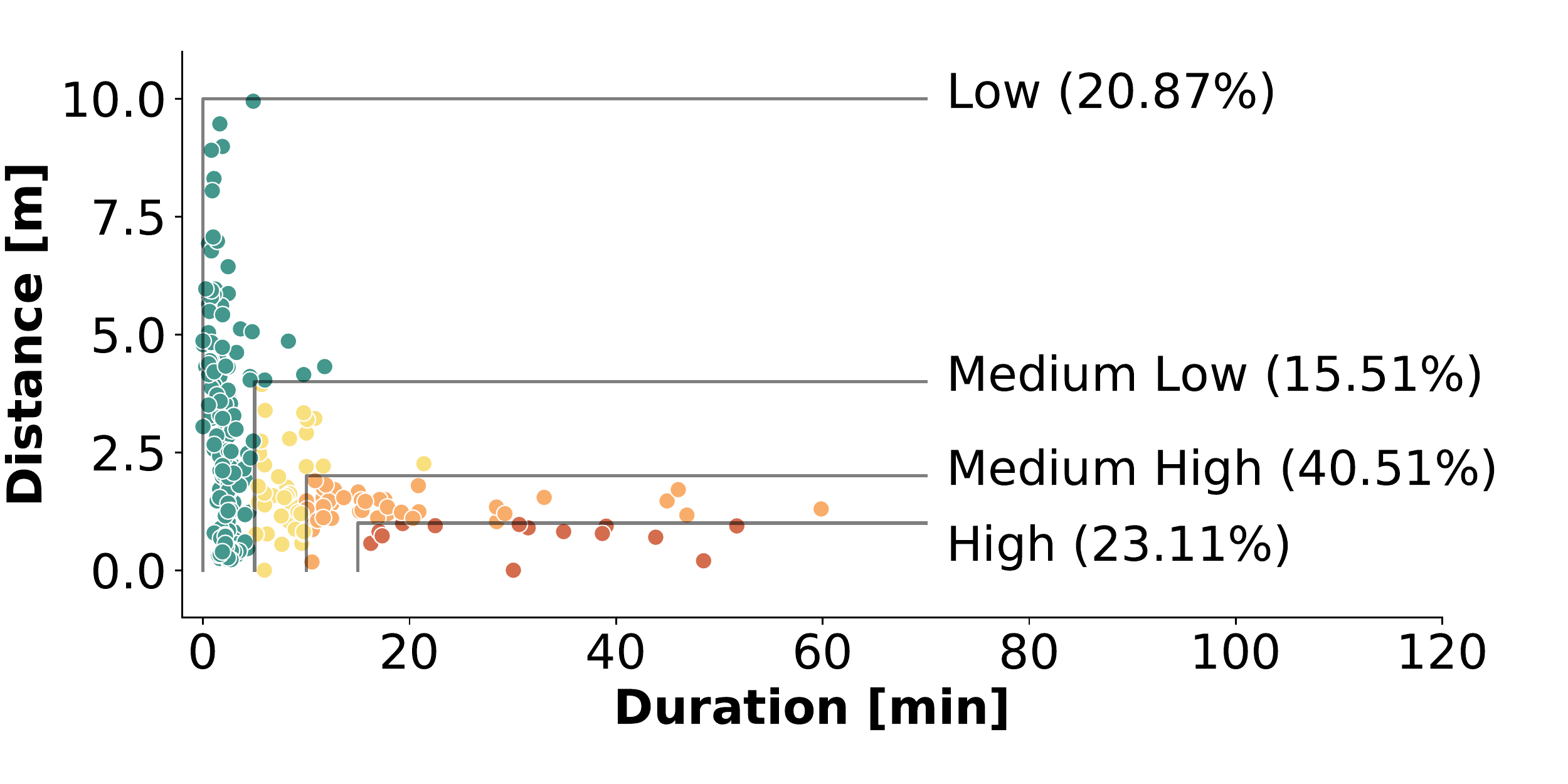}
\caption{
  Maximal duration of exposure.}\label{fig:risk_unique_dur}
\end{subfigure}

\caption{ {Unique contacts and risk levels.} Contacts from
  \idandalo, aggregated into a single point (one per device pair)
  according to different criteria.
}
\label{fig:risk_unique_andalo}
\end{figure}

are due to a small group of study participants or if they are common to the whole sample of participants. To see this, we examine the behavior for pairs of 
individuals in Figure~\ref{fig:risk_unique_andalo} for the  \idandalo summer camp. 
summer camp. This unique point can represent the mean duration and proximity over all interactions of the couple (panel a), the interaction with maximum 
duration (panel b), or the interaction with minimal distance (panel c). From panel a
contacts with minimal spatial distance (subfigure c). However if we focus on the contacts with maximal exposure duration (panel b) we see that $24\%$ of the 
pairs of individuals are involved in very high risk interactions. From this, we conclude that the risk of contagion is quite homogeneously distributed among the 
different pairs of individuals, except for some for whom the longest interactions are also the most dangerous ones.

In the representation in Figure~\ref{fig:risk_all}, each dot
represents a single contact between two individuals, but it ignores
information about the corresponding IDs. Therefore, it is possible
that the analyzed population has heterogeneous behaviors, e.g., with
only a few participants involved in more risky close proximity
interactions and the majority of individuals interacting safely, or
vice-versa. To understand how the risk is distributed among the summer
camp population
we consider three additional views, shown in Figure~\ref{fig:risk_unique_andalo},
where we examine the behavior for pairs of individuals. We report only
the case of \idandalo, since the other camps yielded analogous
results.  First, in Figure~\ref{fig:risk_unique_mean}, we compute for
each pair the average distance and duration across all the contacts,
resulting in a single dot per pair. We observe that each pair
interacts, on average, in low-risk social interactions.  A similar
result is observed in Figure~\ref{fig:risk_unique_dist}, where we
select the single contact per pair with the smallest proximity
distance.  Finally, Figure~\ref{fig:risk_unique_dur} shows the single
contact per pair with the longest duration. Here, we see that $\sim$23\%
of the pairs of individuals are involved in very high-risk
interactions. From this, we conclude that the risk of contagion is
distributed quite homogeneously among the different pairs of
individuals, except for some for which the longest interactions are
also the most dangerous ones.

These graphical representations give a first, general idea of the contact risk levels and offer an understanding of how the risk is distributed among the 
individuals. 
We note that these analyses depend on our definition of contact and, particularly, on the thresholds defined in Section \ref{sec:contact_def}. 

In addition, the proposed contact definition allows us to perform two types of meta-analysis based on the risk levels related to: $(i)$ group dynamics (e.g., 
CPIs among group members, among members of different groups, educator-child interactions, child-child interactions), and $(ii)$ the type of educational and 
recreational activities planned during the summer camp.

As described in Section~\ref{sec:data_description}, each summer camp
setting organized participants in small groups and in specific roles
(educator, child). Groups are intended to keep participants separated
into disjoint bubbles~\cite{block2020social,leng2020effectiveness} so
that any contagion event would remain localized. On the other
hand, roles reflect the internal organization of the summer camps, where
both users (children) and educators (adults) were present.  The
results are graphically reported in Figure~\ref{fig:bar_group}, where
the colored bars show the relative percentages of contacts for each
risk level that can be attributed to child-child, educator-child, and
educator-educator interactions, respectively. Moreover, these can be
divided into interactions involving two people belonging to the same
group (``intra-group'') and those bridging two different groups
(``inter-group''). Instead, the large grey bars in the background report
the total percentages of contacts for each specific type
of interaction, independently on the associated risk. To facilitate
the quantitative comparison of the results,
Table~\ref{tab:groups_stats} reports, for each summer camp, the number
and the total duration of the contacts in the six groups.

When a contact occurs between two members of the same group, we assign to it the activity being performed at that moment by that group. In this way, we add 
another layer of analysis that allows us to study the relationship between the activity type, the number, and the contagion risk level of the contacts. The 
results are shown in Figure~\ref{fig:bar_act}, where we report four bars for each activity, representing the four risk levels. The height of the bars represents 
the sum of the duration of all contacts during each activity divided by the total duration of the activity. Hence, each bar reports the risk per unit time of 
each activity. This normalization allows comparison across the different activities, independent of their duration. The percentages show the fraction of contact 
time within each risk level, for each activity.

\section{Discussion}  
We already observed that in all summer camps there is a non-negligible
percentage of contacts at high risk of contagion and that this is in
general not due to some specific individuals or couples of individuals
but the risk is quite homogeneously distributed among all the
participants
(Figure~\ref{fig:risk_all}--\ref{fig:risk_unique_andalo}).  We now
discuss more in detail the results and their implications.

\begin{table}[!t]
\caption{
Summary of the number and duration of the contacts in the three camps according to the social bubble strategy. For each camp \idandalo, \idpovokids, and 
\idpovoteens, we report for the different bubbles the total time of contact and the number of contacts organized by the role of the participants.
}
\label{tab:groups_stats}
\begin{tabular}{p{12mm}p{14mm}|p{12mm}p{12mm}p{12mm}|p{12mm}p{12mm}p{12mm}}
&& \multicolumn{3}{c|}{Intra-group}&  \multicolumn{3}{c}{Inter-group} \\
\hline
&& child child & child $\;\;$ educator & educator educator & child child & child $\;\;$ educator & educator educator \\
\hline
\idandalo&Time [min] & 10362.40 & 2285.28 & 77.82 & 462.12 & 181.77 & 11.75 \\
&Number & 5484 & 1297 & 52 & 295 & 121 & 10 \\
&&&&&&&\\
\idpovokids&Time [min]& 12075.02 & 5250.60 & 290.32 & 538.95 & 383.93 & 72.33 \\
&Number& 4388 & 2064 & 145 & 341 & 195 & 49 \\
&&&&&&&\\
\idpovoteens&Time [min]& 7732.58 & 2229.58 & 4.22 & - & - & - \\
&Number& 2004 & 613 &  2 &  - &  - &  - \\
\hline
\end{tabular}
\end{table}

\subsection{Social Bubbles and Roles}

\begin{figure}[!p]
\centering
\captionsetup{type=figure}

\begin{subfigure}{.7\textwidth}
  \captionsetup{type=figure}
  \centering  
\hspace*{-8mm}\includegraphics[height=.23\textheight]{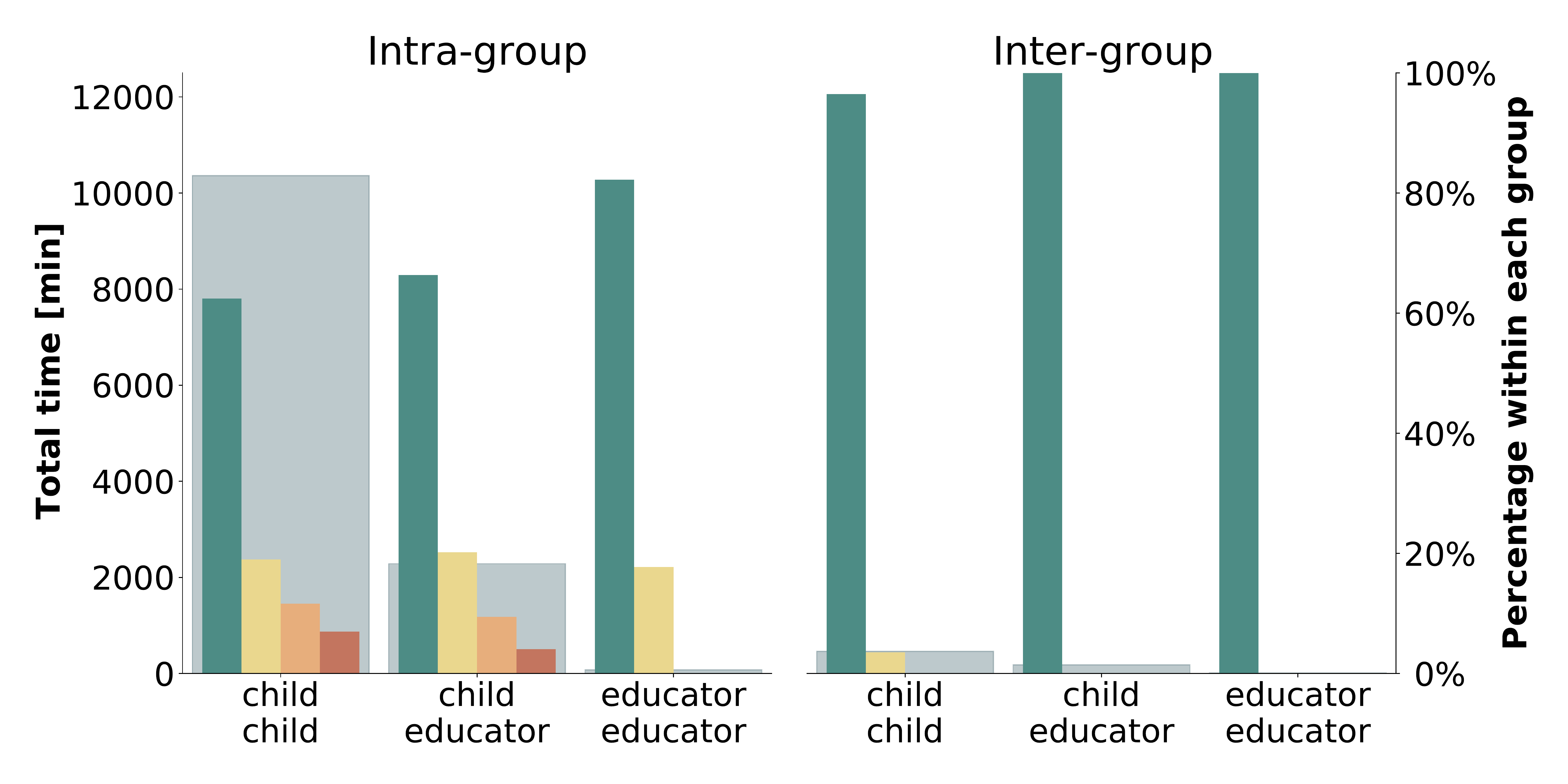}
\caption{\idandalo}\label{fig:bar_group_andalo}
\end{subfigure}

\begin{subfigure}{.7\textwidth}
\captionsetup{type=figure}
\centering
\hspace*{-8mm}\includegraphics[height=.23\textheight]{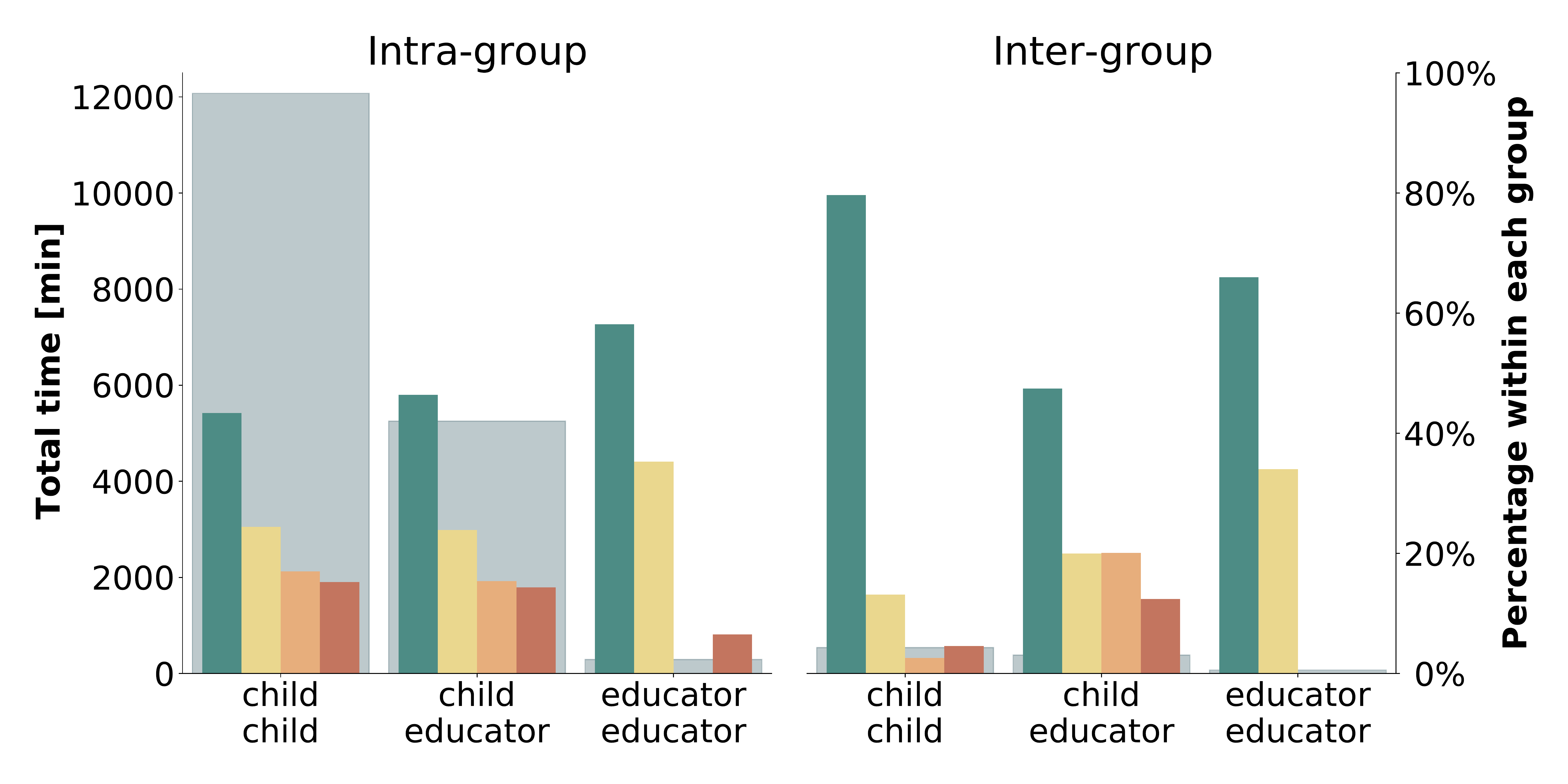}
\caption{\idpovokids}\label{fig:bar_group_povokids}
\end{subfigure}

\begin{subfigure}{.7\textwidth}
  \captionsetup{type=figure}
\centering
\includegraphics[height=.23\textheight]{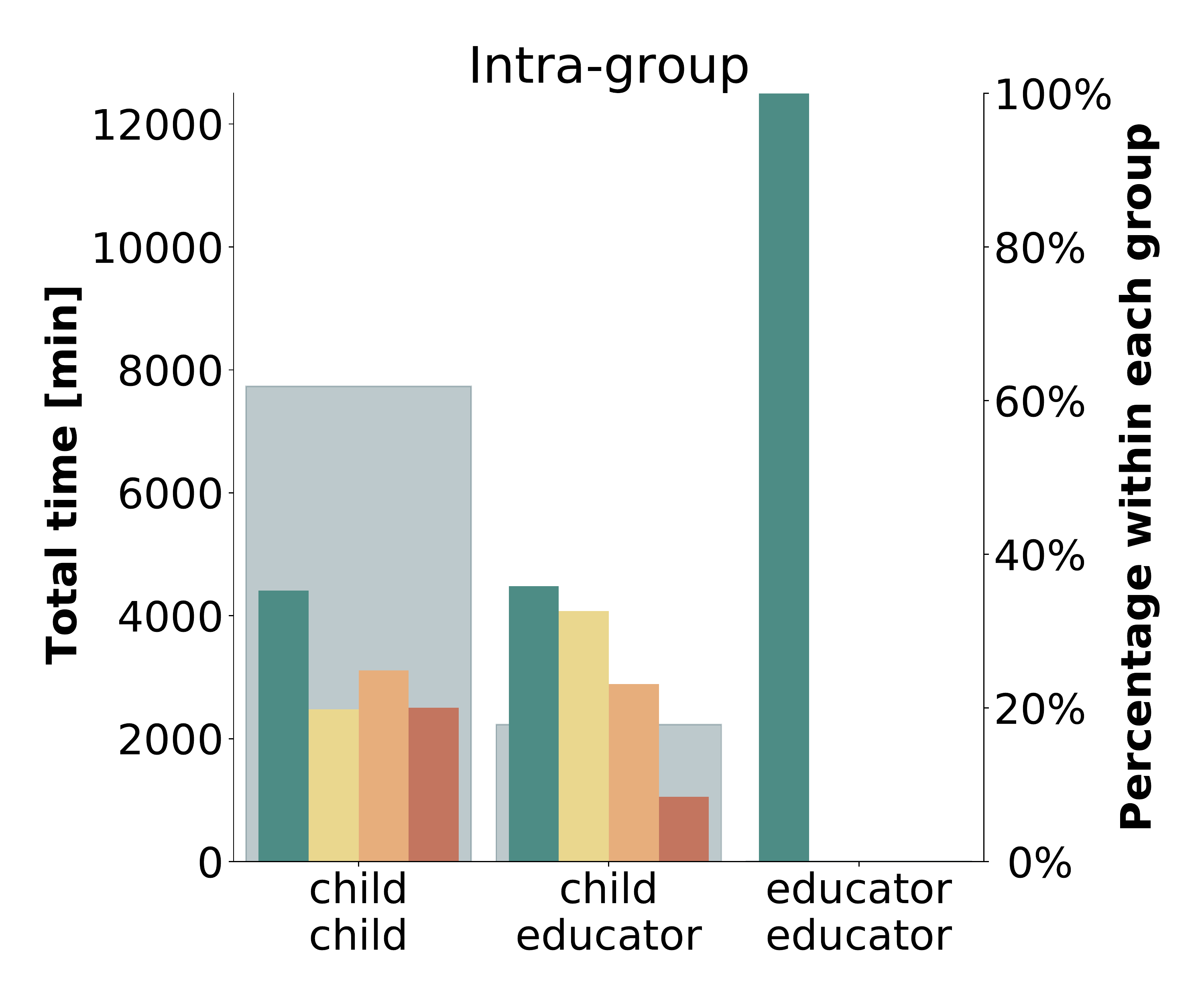}
\caption{\idpovoteens}\label{fig:bar_group_povoteens}
\end{subfigure}
\caption{ {Social bubble policy and roles.} Distribution of
  risk levels by group and type of interaction for each summer
  camp. The color bars, which refer to the right-hand scale, report
  the percentage of time of contact within each risk level.  The grey
  background bars, which refer to the left-hand scale, report the
  total time of contact for each of the six categories.}
\label{fig:bar_group}
\end{figure}

To analyze the effectiveness of the social bubble policies, we look at
Figure~\ref{fig:bar_group},
which reports the percentages of contacts taking place inter- and
intra- groups and between children and children, educator and
educator, and educator and children for the three summer camps. Note
that in \idpovoteens there was only a single group.  We observe, as
expected, that intra-group contacts are more numerous, but they are
also interpreted as less risky since they are foreseen and permitted
within the social bubble policies. On the other hand, inter-group
contacts happen across different groups and are generally more risky;
however, their limited number is a good indication of the
effectiveness of the application of the social bubble policies. The
collected data thus confirm that in case of an epidemic spreading in
these settings, most of the possible contagions would likely be
restricted to a single group, and transmission to other groups
would be avoided or limited.  Focusing on the interactions within each
group, we observe that the highest percentages of contacts with high
or medium-high risk of contagion involve children (i.e.,
children-children or educator-children CPIs), while the educators tend
to have low-risk interactions among them.

\subsection{Activity Type}

\begin{figure}[!p]
\centering
\captionsetup{type=figure}

\begin{subfigure}{.85\textwidth}
\captionsetup{type=figure}
\centering\includegraphics[height=.27\textheight]{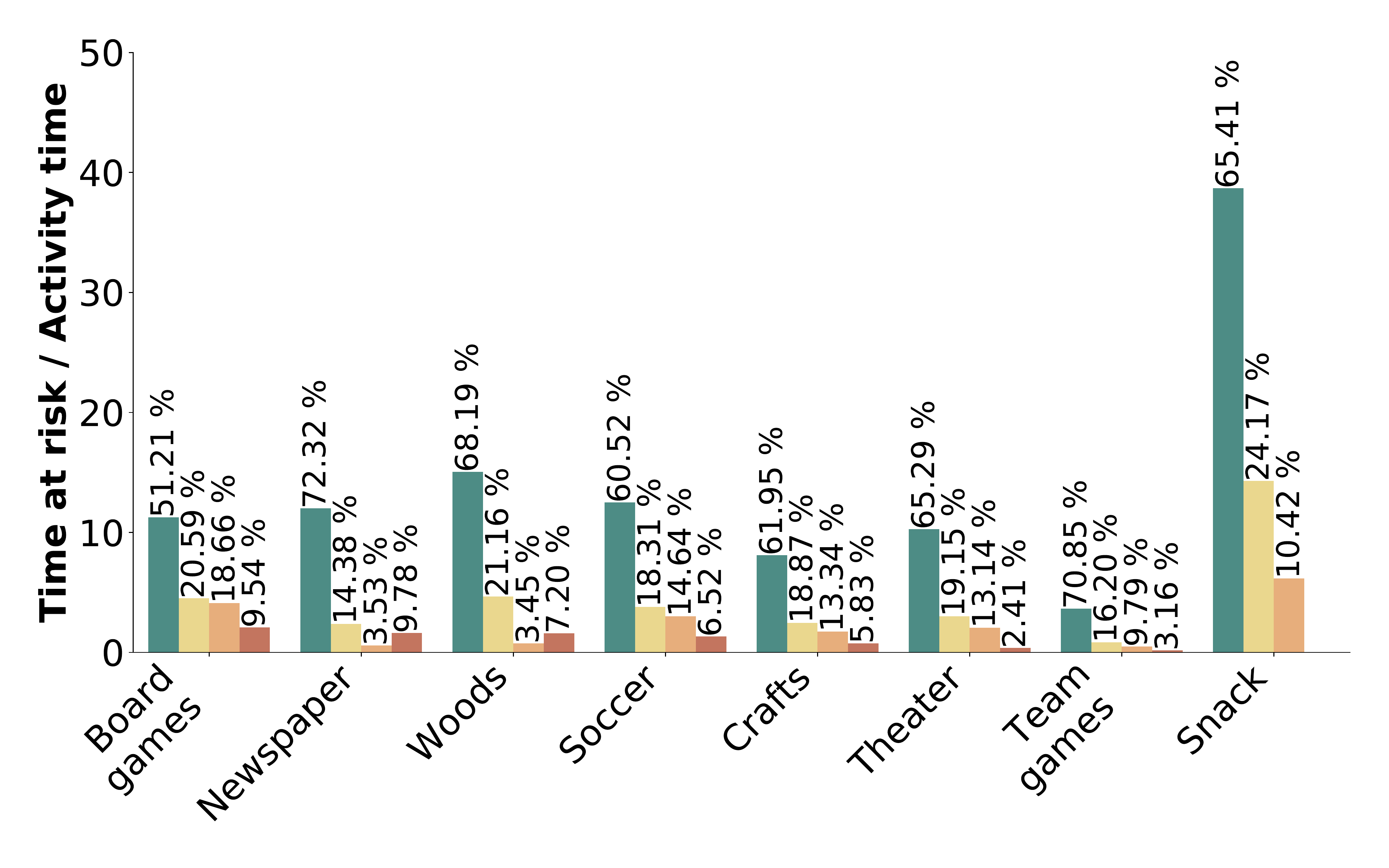}
\caption{\idandalo}\label{fig:bar_act_andalo}
\end{subfigure}

\begin{subfigure}{.85\textwidth}
\centering
\captionsetup{type=figure}
\includegraphics[height=.27\textheight]{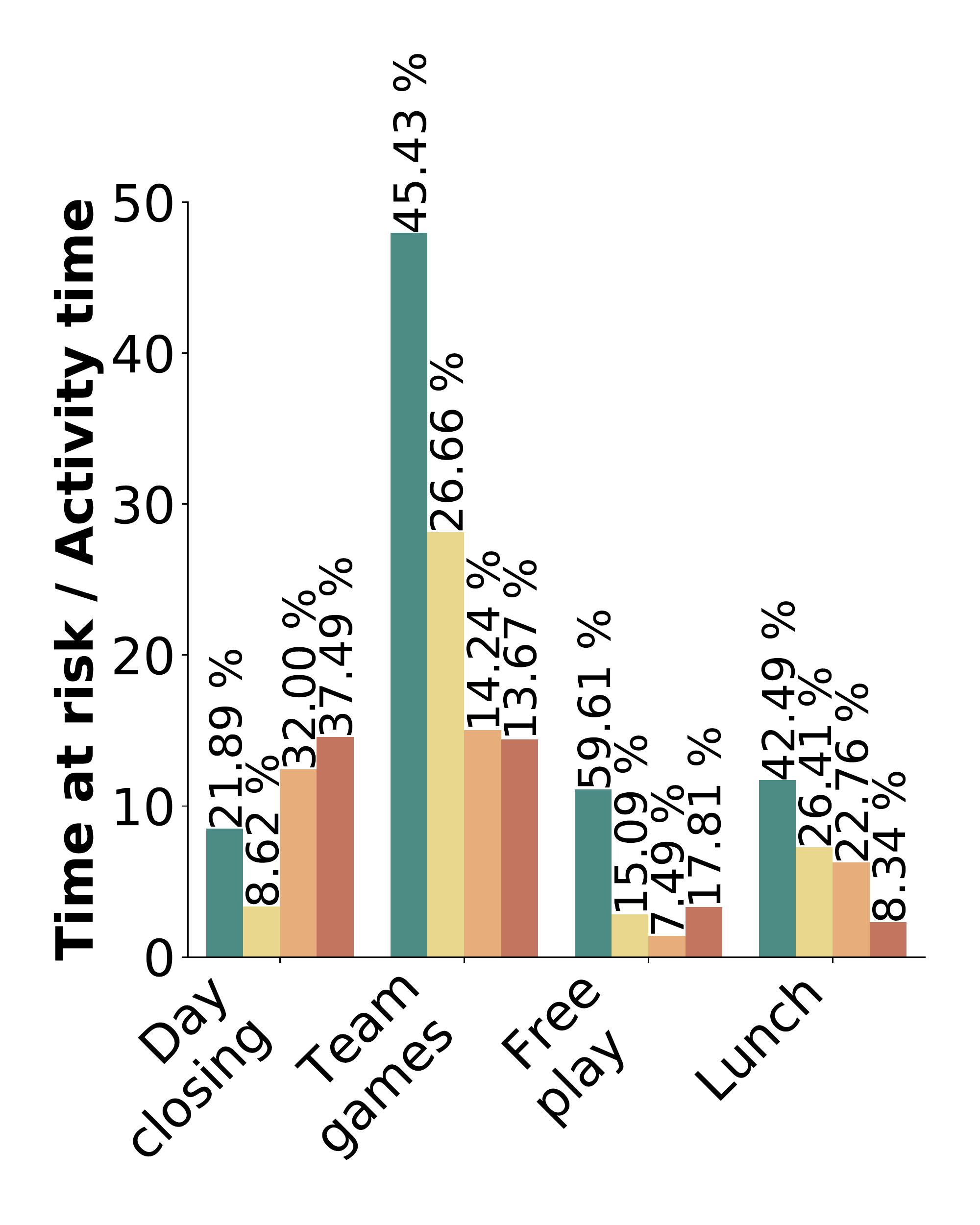}
\caption{\idpovokids}\label{fig:bar_act_povokids}
\end{subfigure}

\begin{subfigure}{.85\textwidth}
\centering
\captionsetup{type=figure}
\includegraphics[height=.27\textheight]{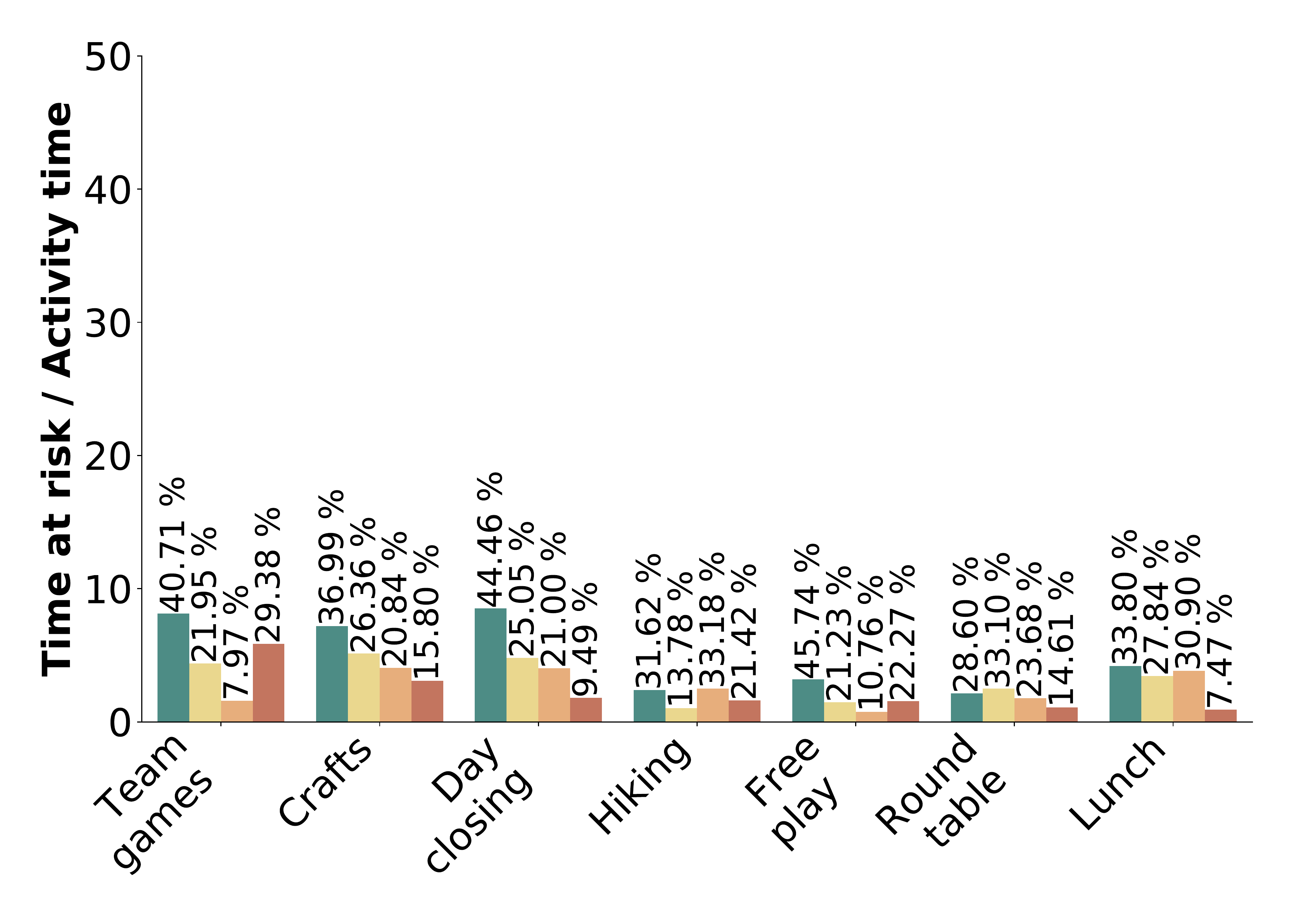}
\caption{\idpovoteens}\label{fig:bar_act_povoteens}
\end{subfigure}

\caption{ {Activities and risk levels}. The figure shows the distribution of the risk levels by activity, sorted according to a decreasing percentage of 
High-risk contacts for \idandalo (Figure~\ref{fig:bar_act_andalo}),  \idpovokids age range 6-11 (Figure~\ref{fig:bar_act_povokids}) and \idpovoteens 11-14 
(Figure~\ref{fig:bar_act_povoteens}).
The percentages show the fraction of contact time within each risk level, for each activity.
}
\label{fig:bar_act}
\end{figure}

For summer camp \idandalo shown in Figure~\ref{fig:bar_act_andalo}, it is evident that the activity involving the highest number of interactions per unit time 
is ``snack''; however, it is also the only activity where none of the CPIs was at high risk. This is actually by design as the activity duration is less than 
15~min (Table~\ref{tab:activities}), which is the minimum duration required to mark a contact as high risk (Table~\ref{tab:risk}).
We observe a similar finding in the other two data sets, \idpovokids and \idpovoteens (Figs.~\ref{fig:bar_act_povokids} and \ref{fig:bar_act_povoteens}), where 
``lunch'' is the activity with the fewest risky contacts. This is probably because, during meal times, the children were not wearing their face masks; thus, the 
educators were paying more attention to the compliance to physical distancing rules. Moreover, the children were seated during lunch, so there was a reduced 
probability of accidental CPIs. 

Other low-risk activities in \idandalo were ``crafts'', ``theater'' and ``team games'', all meticulously organized activities where the educators established 
precise rules for physical distancing to avoid CPIs. The risk rises instead with ``soccer'' and ``woods'', where no precise rules were established, and the 
children were free to move in a large space. Moreover, these activities took place outdoor, and there is evidence for a reduced transmission risk during outdoor 
activities as compared to indoor ones~\cite{lan2020,leclerc2020,bulfone2021,rowe2021}. 

The riskiest activities,  still with a limited total duration
of high-risk close proximity contacts, are represented by ``newspaper'' and ``board games'', two indoor activities with specific constraints: the first 
consisted of collaborating in pairs in front of a computer, working on the summer camp's newspaper, and the second one consisted of playing board games around a 
table. Since the activities required being close to each other watching the same screen or table, the physical distance clearly could not be very large. 
However, it is worth highlighting that children wore face masks during the activities, thus reducing the transmission risks~\cite{li2020,brooks2021,rader2021}.

Moving to \idpovokids, a different summer camp with a different
organization (Figure~\ref{fig:bar_act_povokids}), we observe a high
number of contacts during the activity ``team games'', even if most of
these contacts are at low risk of contagion. Interestingly, in this
summer camp the organized games imply many more contacts per unit time
with respect to ``free play''. However, the activity with the highest
percentage of high-risk CPIs is ``day closing'', which was the final
part of the day, when children were waiting for pick up and
entertained themselves by playing table tennis or table football, in
rather unstructured way.

An additional and final scenario can be observed in \idpovoteens, showing different typical behaviors, possibly due to a higher age range of the participants, 
namely 11-14 years old, and different adherence to physical distancing rules. Figure~\ref{fig:bar_act_povoteens} shows a general lowering of the time spent 
interacting with each other and, at the same time, a higher percentage of high-risk CPIs. Differently from \idandalo but similarly to \idpovokids, we observe 
that the activities with the highest risk are exactly the most organized ones: ``team games'' and ``craft'', followed by the ones where children were more free 
to move around: ``day closing'', ``hiking'', and ``free play''. The activities that provide less high-risk CPIs are instead ``round table'' and ``lunch'', where 
participants were sitting to talk or eat, all together but keeping a well-defined physical distance from one another.

All together, this analysis of the activities shows the different ways in which different settings have been addressed. In particular, it seems that the 
combination of mask-wearing in the close-interaction static activities and a precise organization of the dynamic activities results into an overall effective 
strategy to contain the risk.

\subsection{Limitations}

As with any experimental data collection, we acknowledge the
limitations of our study.  First, the gathered data sets are limited
in time by the duration of the summer camps (one week, and half or
whole days only) and by the number of participants (61 individuals in
total). While the high temporal and spatial resolution enabled by
Janus allow interesting analyses, the sample size and length limits
make it impractical to simulate an epidemic spreading model based on
this population.  Further, all the summer camps were located in the
Trentino area, and do not necessarily directly translate to other
cities, regions, or countries, perhaps with different distancing
rules.

Finally, a comparison to similar studies in the summer camp setting is
not possible, as none are available in the literature. Moreover, we do
not have hard ground truth to compare against; this would have
required either cameras or manual annotations, which would have
greatly interfered with the children privacy and the camps'
activities.  Nevertheless, the results and findings we outlined have
been shared with the educators, who confirmed them based on their
knowledge and recollection of the activity organization, and
the observed general behavior of the children and educators.

Despite these limitations, we reassert that the data collected by the
Janus devices is, to the best of our knowledge, the only example of
physical distance data for child interactions with high
spatio-temporal resolution collected during the COVID-19 pandemic.

\section{Conclusion} 
Tracking and measuring CPIs in a real setting is a challenging task that, however, plays a crucial role in understanding the dynamics of social interactions 
during the pandemic and their effect on the spread of the disease. 

This work shows that the Janus system is well-suited to
provide high temporal and spatial resolution data to capture CPIs
in complex settings like summer camps. Similar observations would have
been impossible to obtain with either BLE or UWB alone.

In particular, we have analyzed three summer camps' daily activities and social interactions in the Autonomous Province of Trento (Italy). The captured CPIs 
allowed us to derive several key insights into the duration and proximity patterns characterizing the child-child and the educator-child interactions. 

Specifically, we verified the effectiveness of the social bubble strategy, which is easy to implement in the summer camp setting and offers an effective 
mechanism to balance control of the epidemic against light restrictions on the children during educational and recreational experiences.

Moreover, we analyzed the risk levels of a series of activities performed during the summer camps. We obtained key information into their safety in terms of 
number of contacts, duration of the contacts, and level of contagion risk. When combined with other metadata such as the location (indoor vs. outdoor) and the 
possibility to adopt personal protective equipment (i.e., face masks), this information may result in actionable policies to design safer environments for 
interactions among children in the summer camp setting but also at schools.

in the hope of advancing the open sharing and collection of face-to-face interaction data in complex social settings. 

\section*{Availability of data and materials}
The datasets generated and analysed during the current study are available from the authors on reasonable request. Please contact Bruno Lepri (lepri@fbk.eu) or 
Amy L. Murphy (murphy@fbk.eu).

\section*{Competing interests}
The authors declare that they have no competing interests.

\section*{Author's contributions}
Conceived the study and data collection: BL, EF, EL, and ALM. Designed and developed Janus: EL, TI, DM, GPP, ALM. Designed and performed the experiments: EL, 
GC, GS, BL, EF, ALM. Analyzed and evaluated the results: EL, GC, GS, GPP, BL, EF, ALM. Wrote the paper: EL, GC, GS, GPP, BL, EF, ALM. All the authors read, 
reviewed and approved the final manuscript.

 \section*{Acknowledgments}
 The authors would like to thank the Agency for Family, Birth and Youth Policies of the Autonomous Province of Trento, the two social cooperatives that 
organized the summer camps, the educators, the children (study participants) and their parents for their essential participation that made this project 
possible.

\section*{Funding}

The development of Janus was partially funded by the VRT Foundation
(Fondazione per la Valorizzazione della Ricerca Trentina). This work
was partially funded by EIT Digital (ProxyAware project, Activity
20666) and by the Italian government (NG-UWB project, MIUR PRIN 2017).

\bibliography{biblio}
\bibliographystyle{bmc-mathphys}

\appendix
\section{Pre-processing of the data}\label{si:pre_processing}

Prior to analysis, the data collected during each summer camp were cleaned of spurious samples recorded by the devices. We describe the process here and report 
a summary of the collected data for each setting.

The Janus devices do not have an on/off switch, and as a result, are active 24 hours per day, not only when the summer camps are in session. Although we used 
the inhibitor device to limit the measurements taken after the daily close of the summer camp, some additional measurements are still stored.

For example, if the BLE signal to the inhibitor was weak, the devices may have been briefly activated. Additionally, the inhibitor node was often disabled 
several minutes before children arrival and devices distribution, resulting in measurements among the devices still on the storage bench. Finally, some children 
were absent for entire days or arrived late while their device was still taking measurements.

Identifying all these cases was a largely manual effort based on information from the educators about absences and observations in the data itself. For example, 
when a sequence of constant distance measurements is seen at the beginning of the day, it is likely that the devices are still in storage, as children are 
rarely so still.  
The data cleaning step filters all these spurious measures. Table \ref{tab:datasets_raw}shows for each summer camp the data collection time frame, the number of 
unique participants that have been involved, the number of overall measures, and the number of measures after the filtering step.

\begin{figure}[!t]
    \centering
    \captionsetup{type=figure}
    \includegraphics[width=.9\textwidth]{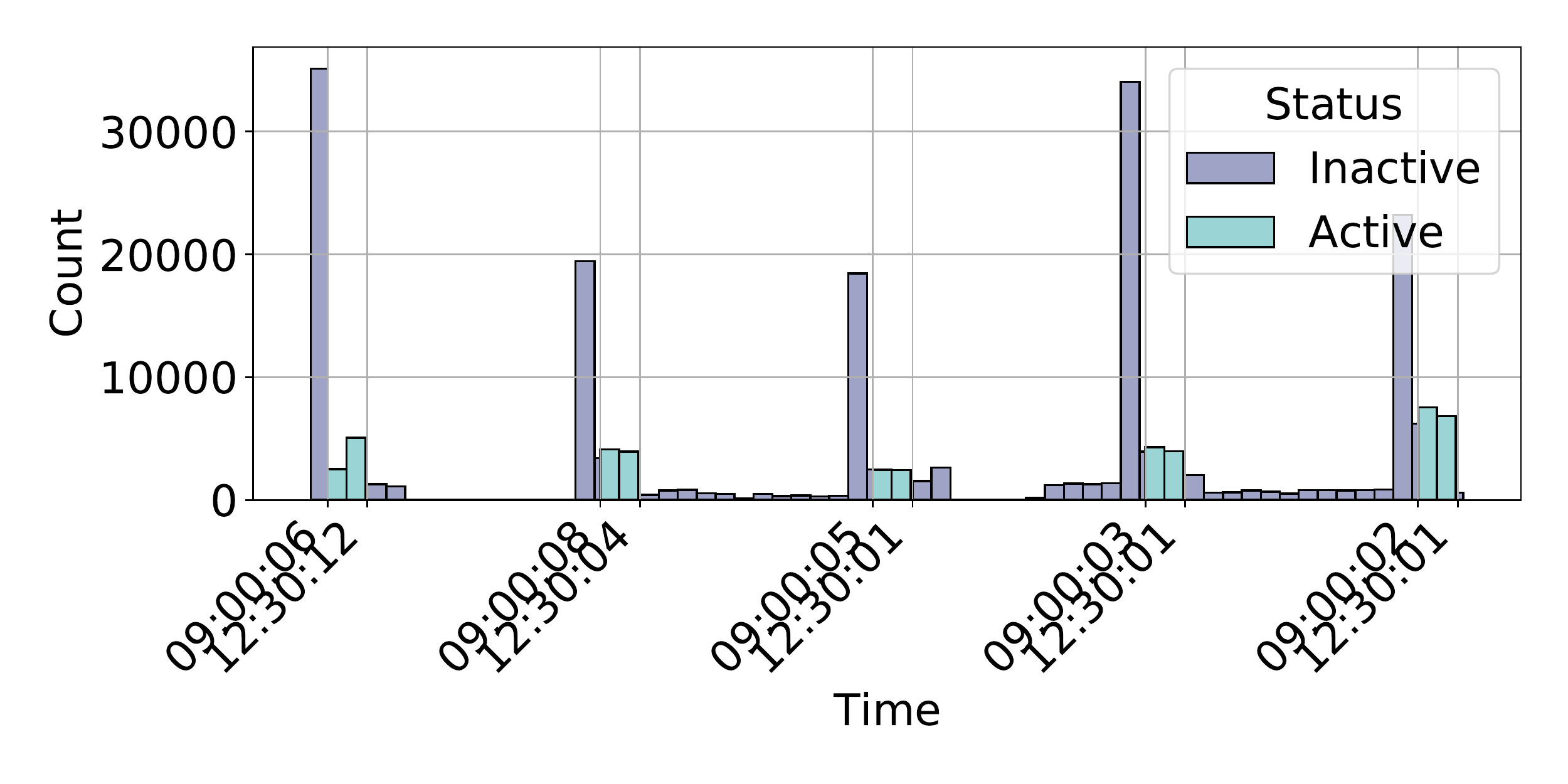}
    \caption{ {Filtering of the spurious measures.} Distribution of the measurements over the entire sampling period, either with \textit{Active} or \textit{Inactive} status for \idandalo. 
    }\label{fig:preprocessing_all_data}
\end{figure}

Figure~\ref{fig:preprocessing_all_data} shows the distribution of the entire measurement set for \idandalo. The time intervals during which the activities took 
place ({\em Active}) are separated from the time between the activities ({\em Inactive}). The peaks of data close to the morning camp start time correspond to 
the phase when the inhibitor node is off, but the devices have not yet been distributed to the children. In this case, all devices are immobile, near one 
another on a bench (Figure~\ref{fig:andalo}) and thus save many distance measurements.

\begin{table}[!t]
\caption{Statistics of the raw data sets, including the number of measures before and after the pre-processing step.
}
\label{tab:datasets_raw}
\begin{tabular}{p{25mm}p{15mm}p{15mm}p{15mm}p{15mm}p{15mm}}
\hline
ID & Initial day & Final day & Unique users &  Raw $\quad$ measures & Filtered measures\\
\hline
\idandalo & 2020-08-17 & 2020-08-21 & 24 & 222222 &\phantom{1}48739\\ 
\idpovokids + \idpovoteens  &2020-08-24 & 2020-08-30 & 25 & 213219 & 146576 \\ 
\hline
\end{tabular}
\end{table}

\end{document}